\documentclass[preprint,showpacs,preprintnumbers,amsmath,amssymb]{revtex4}

\usepackage{graphicx}
\usepackage{dcolumn}
\usepackage{bm}

\begin{document}


\title{Synchronization on fast and slow dynamics in drive-response systems}

\author{Liang Wu}
\email{liangwu@suda.edu.cn}
\author{Shiqun Zhu}
\altaffiliation{corresponding author}
\email{szhu@suda.edu.cn}
\author{Juan Li}

\affiliation{China Center of Advanced Science and Technology
(World Laboratory), P. O. Box 8730, Beijing 100080, PRC}

\affiliation{School of Physical Science and Technology, Suzhou
University, Suzhou, Jiangsu 215006, People's Republic of
China\footnote{Mailing address} }


\begin{abstract}
Two types of synchronization, Achronal Synchronization and
Isochronous synchronization are investigated numerically when
unidirectionally coupled laser systems are considered both on fast
and slow dynamics by studying the correlation function. Although
the synchronization behaviors are found to be separated on
relatively fast dynamics of chaotic fluctuations while blended
together on slow dynamics of dropout events, their main features
revealed are in a good agreement, which most probably suggests
that the observation based on detections on slow dynamics is of
importance when it is the only choice as detection on fast
dynamics is usually extremely hard to be done.

\end{abstract}

\pacs{05.45.Xt, 42.55.Px, 42.65.Sf}
\maketitle
\section{Introduction}
Synchronization between coupled chaotic oscillators is a
fundamental phenomenon widely observed in science and
nature\cite{Pecora90PRL}. It has various applications, especially
in secure communication where it can be used to enhance the
privacy\cite{VanWiggeren98SCI}. For two coupled identical
oscillators, Complete Synchronization (CS) may occur between
them\cite{Tang03PRL,Masoller01PRL,Ahlers98PRE,Liu01apl}. However
there are often correlated entrainments of discrete events when
nonidentical ones are considered, meanwhile their fast chaotic
behaviors remain asynchronous, which is often referred to as phase
synchronization (PS)
phenomenon\cite{Rosenblum96PRL,Rosenblum97PRL,Ivanchenko04PRL(1)}.
Typical examples are: (i)neuronal system may produce common
rhythmic bursting as there is coupling between neurons, while its
individual neurons still show asynchronous dynamics of fast
spiking\cite{Elson98PRL,Rulkov01PRL,Dhamala04PRL,Ivanchenko04PRL(2)};
(ii)coupled semiconductor lasers operating in low frequency
fluctuation (LFF) may produce rhythmic dropout events whereas
their fast chaotic dynamics remain largely
different\cite{Wallace01PRA,Wedekind02PRE,Buldu02apl,Heil01PRL,Sivaprakasam01PRL}.
On the other hand, the existing coupling would still introduces
some correlation between the involved oscillators in spite that
their dynamics are seemly completely asynchronous. It is possible
that there is certain functional relations between their outputs,
a situation which is often called generalized synchronization
(GS)\cite{Rulkov95PRE,Kocarev96PRL}, while those explicit
functions prove to be difficultly recognized in many other cases,
the still existing correlation would show itself by some
similarities function, such as correlation function, mutual
information function and so on.\cite{Rosenblum97PRL}

Although a large number of papers about synchronization have been
published in which both fast and slow dynamics are
concerned\cite{Wedekind04Chaos,Dhamala04PRL,Ivanchenko04PRL(2),Elson98PRL},
unfortunately little work has been done  in such framework on
Achronal and Isochronous synchronization which are simultaneously
existing ubiquitously in most drive-response systems.

This two qualitatively different types of synchronization,
Isochronous and Achronal
synchronization\cite{Locquet02PRE,Koryukin02PRE,Murakami02PRA,Locquet02OL,Uchida04PRE,Voss00PRE},
are discussed in this paper when both fast and slow dynamics are
concerned. On one hand the directed coupling from one laser to
another may be strong enough to lead to a locking state phenomenon
and consequently Isochronous Synchronization (IS). On the other
hand, Achronal Synchronization (AS) may also occur, corresponding
to the mathematical solution of equations describing the
subsystems' behaviors. We show that the synchronization behaviors
are clearly separated on fast dynamics, while blended together on
slow dynamics. Moreover most important features about
synchronization can be well reflected by both fast and slow
dynamics, which most probably suggests the observation on slow
dynamics is of importance when it is the only choice as detection
on fast chaotic dynamics is usually technically hard to be done.

\section{Synchronization States}
The following equations are used to describe a typical
drive-response (directed coupled) system:
\begin{eqnarray}
\dot{X}_t=f(X_t,\gamma X_{t-\tau})\\
\dot{Y}_t=f(Y_t,\eta_{12} X_{t-\tau_{12}})
\end{eqnarray}
where $\gamma$ represents the feedback rate in drive-subsystem(X),
$\eta_{12}$ is the coupling strength from drive-subsystem(X) to
response-subsystem(Y). $\tau$ is the delay involved in feedback,
and $\tau_{12}$ corresponds to the retardation time for the
coupling from X to Y. Two basic types of synchronization probably
occurring in such systems are:
\begin{eqnarray}
Isochronous \ Synchronization &:& Y_t \sim X_{t-\tau_{12}}\\
Achronal \ Synchronization &:& Y_t \sim X_{t+(\tau-\tau_{12})}
\end{eqnarray}

For the reason that synchronization is a universal nonlinear
phenomenon, and its main features are typically independent of
particular properties of a model, a system consisting of two
semiconductor lasers in drive-response configuration as an example
is considered in this paper. One laser acts as the drive-subsystem
(called Master) and it is driven into low-frequency-fluctuation
(LFF) operating regime by being subjected to a feedback introduced
by a mirror. In addition, a part of its output is injected into
another laser acting as the response-subsystem (called Slave).
This injection drives the Slave laser into LFF regime as well as
introduces a coupling directed from Master to Slave. The two
lasers' behaviors can be described and numerically simulated by LK
model\cite{Lang80}.

\begin{figure}
\includegraphics{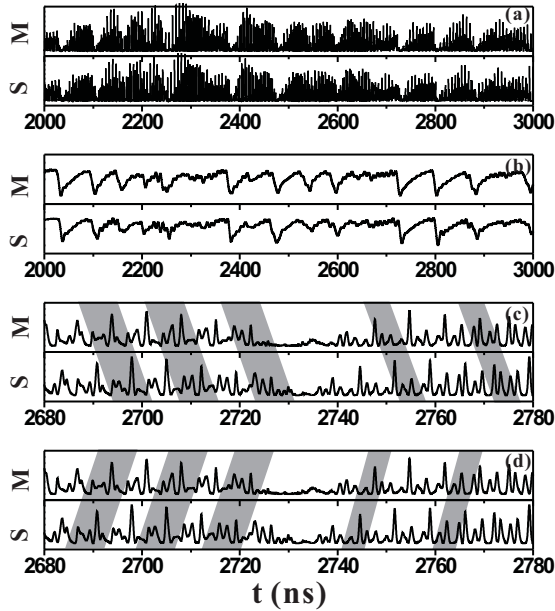}
\caption{\label{fig:epsart} Outputs of two unidirectionally
coupled semiconductor lasers simulated by LK model. chaotic
low-frequency-fluctuation behaviors of two correlated lasers from
2000-3000 nanosecond in (a), M presenting Master laser and S Slave
laser.  Their corresponding low-pass-filtered time series in (b)
where dropout events are highlighted to show the correlation on
slow dynamics. a relatively short period of time series,
emphasizing Isochronous synchronization in (c) and Achronal
synchronization in (d) on fast dynamics.}
\end{figure}

It is found numerically that synchronization phenomenon is
occurring on both fast and slow dynamics. Fig. 1(a) plots chaotic
outputs of Master and Slave lasers both operating in
low-frequency-fluctuation regime from 2000 nanosecond to 3000
nanosecond where it is seen dropout events taking place
rhythmically in two lasers. Those dropouts are further highlighted
in Fig. 1(b) where the low-pass-filtered time series are plotted
and where it is clearly seen there is Phase Synchronization
occurring on slow dynamics in a timescale of dropout events shown
by the easily observable rhythm between the outputs. On the other
hand, a relatively short period of time series ranging from 2680
nanosecond to 2780 nanosecond are shown in Fig. 1(c) as well as in
Fig. 1(d). Some gray parallelograms are added in the plots to make
it clear the similarities between the detailed chaotic outputs of
Master and Slave. The continuously existing similarity
relationship between the lasers' outputs in shadowed as well as
unshadowed areas lasting for a relatively long period of time
together lead to the emergence of the synchronization on fast
dynamics in a time scale of seemly random fluctuations.

We notice the delay in feedback is set $\tau=7ns$ in our
calculation, and the coupling retardation time $\tau_{12}=4ns$.
Focus has been put upon Isochronous Synchronization in Fig. 1(c)
where the output of Master laser is ahead of that of Slave laser
by around 4 nanoseconds, satisfying Eq. (3); while Fig. 1(d)
stresses Achronal Synchronization with the output of Master laser
lagging behind that of the Slave laser by about 3 nanoseconds,
satisfying Eq. (4), although the same piece of time series is
presented in Fig. 1(3) and (4). Therefore the two types of
synchronization behaviors are simultaneously occurring .

\section{Comparison between Fast and Slow Dynamics}

The qualities of synchronization behaviors are usually measured by
correlation function as a function of the shift time ($\tau_s$).
\begin{eqnarray}
C(\tau_s)=\frac{\langle[I_m(t+\tau_s)-\langle
I_m\rangle][I_s(t)-\langle I_s\rangle]\rangle}{\{\langle
[I_m(t)-\langle I_m\rangle]^2\rangle\langle [I_s(t)-\langle
I_s\rangle]^2\rangle\}^{1/2}}
\end{eqnarray}
Complete synchronization should be indicated by $C=1$, and higher
correlation implying a better synchronization. On the other hand,
poor synchronization would give a relatively low correlation C and
hence a low synchronization quality.

\begin{figure}
\includegraphics{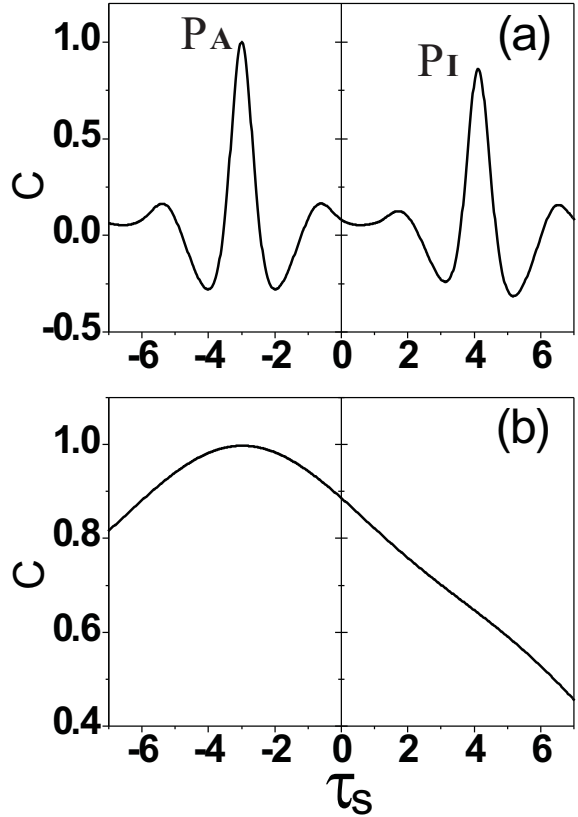}
\caption{\label{fig:epsart} Correlation plots calculated (a) by
original time series to reflect synchronization behaviors on fast
dynamics, (b) by low-pass-filtered time series to show on slow
dynamics. $\gamma=\eta_{12}=7ns^{-1}$}
\end{figure}

Fig. 2 quantitatively presents the synchronization quality of the
two synchronization behaviors. Correlation function plotted in
Fig. 2(a) is calculated by original time series in order to show
the features on fast chaos, meanwhile correlation function by
calculating on low-pass-filtered series is plotted in Fig. 2(b) to
show the features on slow dynamics. There are two peaks emerging
in Fig. 2(a), One ($P_A$) corresponds to Achronal synchronization
and the other($P_I$) to Isochronous Synchronization ($P_I$). The
peak at $\tau_s=4 ns^{-1}$ indicates that a high correlation would
be obtained when the output of Slave laser is shifted forward by
4ns with respect to that of Master laser and satisfying Eq. (3)
and hence corresponds to Isochronous Synchronization. $P_I$
represents the peak value. On the other hand, the peak at
$\tau_s=-3 ns^{-1}$ shows the high correlation obtained when
shifted backward by 3ns and satisfies Eq. (4), therefore Achronal
Synchronization is revealed in this way with its corresponding
peak value $P_A$. In contrast, there is only a hump in Fig. 2(b),
which comes from the fact that the two synchronization behaviors
are blended with each other on slow dynamics.

By comparing the correlation function of fast dynamics in Fig.
2(a) with that of slow dynamics in Fig. 2(b), their difference can
be clearly seen: firstly, the very steep slope on both sides of
the two peaks in Fig. 2(a) prove that the shift time of the
correlation peaks are considerably explicit, $-(\tau-\tau_{12})$
for Achronal Synchronization and $\tau_{12}$ for Isochronous
Synchronization, which most probably indicates that the two peaks
as well as the corresponding synchronization behaviors they
represent are completely separate and uncorrelated on fast
dynamics. In contrast, gentler declination of the correlation from
the hump in Fig. 2(b) with $C_{(\tau_s=0)}=0.88$ much greater than
that in Fig. 2(a) suggests that Achronal and Isochronous
synchronization behaviors are entangled and can not be separated
from each other. As a result, they show themselves by probability
distribution and consequently there is only a hump. Secondly, the
two synchronization behaviors compete with each other. The
competition result is reflected by the comparison which
correlation peak is higher on fast dynamics on fast dynamics,
while on slow dynamics the hump is clearly tilted towards one
side, indicating the synchronization behavior corresponding to
this side is more probably selected over the other by the system.

\begin{figure}
\includegraphics{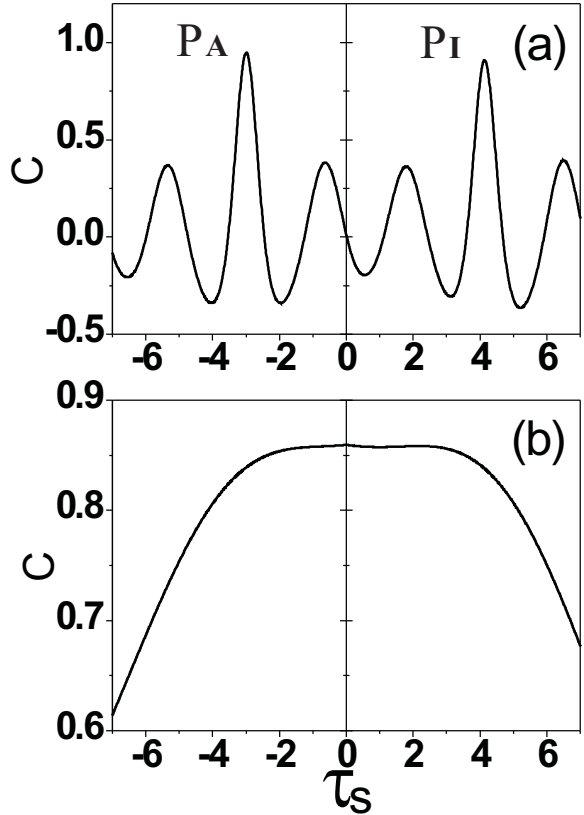}
\caption{\label{fig:epsart}The balance in quality between the two
synchronization behaviors on fast and slow dynamics.}
\end{figure}

Moreover, such blend of the two synchronization behaviors on slow
dynamics will come to its highest level when they are balanced in
quality under a very special condition. In Fig. 3(a) the nearly
equi-altitude correlation peaks on two sides shows the balance in
quality of the behaviors on fast dynamics, on the other hand the
two peaks surprisingly turn into a single correlation plateau when
slow dynamics are concerned in Fig. 3(b). It is seen
$C_{(\tau_s=-3)}= C_{(\tau_s=0)}=C_{(\tau_s=4)}$ on this plateau,
which indicates the hump as shown in Fig. 2(b) would tilt to
neither side, namely neither of the synchronization behaviors are
selected by the system over the other. The correlation peaks
therefore disappear completely, reaching a so-called "free state"
in which there is no certain value of $\tau_s$ by which time
series are shifted to be able to ensure a greater correlation.

Fast and slow dynamics are both capable of showing the main
dynamical features of the occurring synchronization. The following
part will focus on the correspondence between them. Fig. 4 shows
how the result of their competition depends on the coupling
strength. Here the competition result means which one of the two
synchronization behaviors will be more probably selected by the
system over the other. For fast dynamics the result can be shown
by $P_I-P_A$, the difference between the two correaltion
peak-values, see Fig. 4(a). A larger $P_I-P_A$ indicates that
Isochronous Synchronization is more probably selected over
Achronal synchronization, and the system's dynamical behavior, as
a result, is mainly governed by the former. Such domination can be
clearly shown as well by the fact that the similarities in Fig.
1(c) emphasizing Isochronous Synchronization would have been more
recognizable and easily seen than those in Fig. 1(d) which
emphasize Achronal synchronization. In contrary, a minus value of
$P_I-P_A$ would indicate Achronal Synchronization is the one
ruling synchronization behavior of the system .

\begin{figure}
\includegraphics{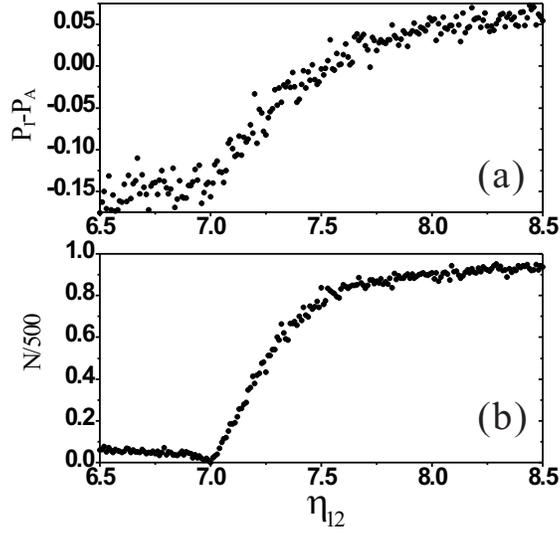}
\caption{\label{fig:epsart}Competition result between the
synchronization behaviors, shown on fast dynamics by $P_I-P_A$ and
on slow dynamics by $N/500$, depending on the coupling strength
$\eta_{12}$.}
\end{figure}

As for slow dynamics, a quantity $N/500$ is able to measure the
competition result by showing the probability that the dropouts in
Master laser are ahead of the corresponding ones in Slave laser.
In the calculation of $N/500$, 500 dropout pairs are identified
firstly by the condition that two dropouts from different lasers
are paired when they are close to each other with the time
interval less than 8 nanoseconds, subsequently they are judged one
by one which laser's output is ahead in time. Finally the value of
N shown in Fig. 4(b) is obtained by counting the number of those
dropout pairs in which Master's dropout occurs ahead of Slave's in
time. The closer $N/500$ is to 1, the more probably Isochronous
Synchronization is selected by the system over Achronal
synchronization, in contrary $N/500<0.5$ means Achronal
Synchronization is more probably selected. Since not outputs in
detail but just the dropout events are concerned in $N/500$, Fig.
4(b) considers the features on slow dynamics.

In Fig. 4 both $P_I-P_A$ and $N/500$ are seen to increase
gradually in the region from 7.0 to 7.6 ns$^{-1}$, suggesting that
the role of ruling system synchronization behavior has been
transferred from Achronal to Isochronous Synchronization. The
transition process slows down when coupling strength is greater
than 7.6 ns$^{-1}$ in both figures. Therefore there is a good
agreement between fast and slow dynamics. The coupling is too weak
to produce continuous synchronous behavior both on fast and slow
dynamics when $\eta_{12}<7.0$.

\begin{figure}
\includegraphics{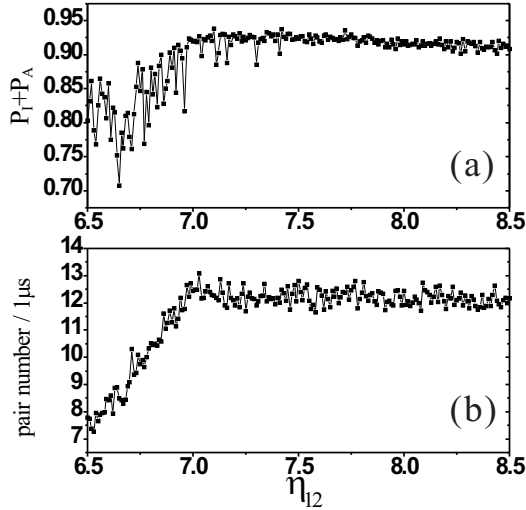}
\caption{\label{fig:epsart} The overall synchronization quality,
shown on fast dynamics by $P_I+P_A$ and on slow dynamics by the
averaged number of dropout pairs every microsecond, depending on
the coupling strength.}
\end{figure}

That the features of synchronization behaviors can be shown by
both fast and slow dynamics is further shown in Fig. 5, where the
dependence of $P_I+P_A$ (a) and the number of dropout pairs per
microsecond (b) on the coupling strength $\eta_{12}$ is plotted.
Since on fast chaos the two synchronization behaviors are shown by
two separated correlation peaks with the peak-values $P_I$ and
$P_A$ and they take place simultaneously, the summation of $P_I$
and $P_A$ seems useful to estimate the overall synchronization
quality. From $\eta_{12}=6.5$ to 7.0ns$^{-1}$, $P_I+P_A$ has being
increased before it comes to its maximum value
$\eta_{12}=7.0ns^{-1}$ around which the overall synchronization
quality, when $\eta_{12}$ is over 7.0ns$^{-1}$, remains almost
unchanged. These changing procedures are capable of being
reflected by slow dynamics as well. In average there are about
12.5 dropouts per microsecond generated in each laser. However
some dropouts of Master laser are not capable of being connected
with dropouts from Slave laser to form dropout pairs when the
coupling is not sufficiently strong. It is expected when the
coupling is enhanced that the resulting correlation is
strengthened and that the forming-pair rate rises steadily till to
its extreme level so that all of the dropouts can find their
partners from the other laser to form pairs with the number of
pairs per microsecond remaining around 12.5 pairs per microsecond.
Through the comparison Fig. 5(a) with Fig. 5(b), it is obvious
that fast and slow dynamics are in a good agreement when they are
used to study the dynamical features of the occurring
synchronization.

\section{Discussion}

Both Fig. 4(a) and (b) show the fact that the improvement of
overall synchronization quality to its full level is a gradual
course. That is to say, the transition of system behavior from
asynchrony to synchrony is not a sudden change, but a progressive
one where the overall synchronization quality is improved little
by little.

The correspondence between fast and slow dynamics is shown not
only when the overall synchronization has been fully established,
i.e. $\eta_{12}>7.0$, but also in the region where it has not yet
when $\eta_{12}<7.0$. For example, Achronal and Isochronous
synchronization are both improved steadily with the increase of
the coupling coupling strength and the former is always better
than the later when $\eta_{12}$ is less than 7.0ns$^{-1}$. These
things can be reflected by the fact that $P_I$ is always less than
$P_A$ in Fig. 4(a) regarding fast dynamics, and that $N/500$
remains less than 0.1 and almost unchanged in Fig. 4(b) regarding
slow dynamics.

\section{Conclusion}
This paper compares the dynamical features of the synchronization
behaviors revealed by fast dynamics with those revealed by slow
dynamics. Although there are differences especially in the way the
synchronization behaviors are shown, most other important features
about synchronization reflected by fast and slow dynamics are in
agreement, which most probably suggests that the observation based
on the detection on slow dynamics is of importance when it is the
only choice as detection on fast dynamics is technically hard to
be done experimentally.

\begin{acknowledgments}
The financial support from the Natural Science Foundation of
Jiangsu Province (Grant No. BK2001138) is gratefully acknowledged.
\end{acknowledgments}
\bibliography{bistable}
\end{document}